\documentclass[prc,twocolumn,floatfix,groupedaddress,nofootinbib,showpacs,preprintnumbers,
amsmath,amssymb,amsfonts,superscriptaddress,widetable] {revtex4}
\usepackage{bm}
\usepackage{mathrsfs}
\usepackage{amssymb}
\usepackage{amsmath}
\usepackage{graphicx}
\usepackage{array}
\def\alphad{$\!\alpha_{\raisebox{-1pt}{\tiny  D}}\,$}

\begin{document}
\title{Has a thick neutron skin in ${}^{208}$Pb been ruled out?}
\author{F.~J. Fattoyev}\email{farrooh.fattoyev@tamuc.edu}
\affiliation{Department of Physics and Astronomy, Texas A\&M
               University-Commerce, Commerce, TX 75429, USA}
\author{J. Piekarewicz}\email{jpiekarewicz@fsu.edu}
\affiliation{Department of Physics, Florida State University,
               Tallahassee, FL 32306, USA}
\date{\today}
\begin{abstract}
 \noindent
The Lead Radius Experiment (PREX) has provided the first
model-independent evidence in favor of a neutron-rich skin in
${}^{208}$Pb. Although the error bars are large, the reported large
central value of 0.33\,fm is particularly intriguing. To test
whether such a thick neutron-skin in ${}^{208}$Pb is already
incompatible with laboratory experiments or astrophysical
observations, we employ relativistic models with neutron-skin
thickness in ${}^{208}$Pb ranging from 0.16 to 0.33 fm to compute
ground state properties of finite nuclei, their collective monopole
and dipole response, and mass-{\sl vs}-radius relations for neutron
stars. No compelling reason was found to rule out models with large
neutron skins in ${}^{208}$Pb from the set of observables considered
in this work.
\end{abstract}
\smallskip
\maketitle

The Lead Radius Experiment (``PREX'') at the Jefferson Laboratory
has provided the first model-independent evidence on the existence
of a neutron-rich skin in ${}^{208}$Pb\,\cite{Abrahamyan:2012gp,
Horowitz:2012tj}.  Relying on the fact that the weak charge of the
neutron is much larger than the coresponding one in the proton, PREX
used parity-violating electron scattering to probe the neutron
distribution of $^{208}$Pb\,\cite{Donnelly:1989qs}. Elastic electron
scattering is particularly advantageous as it provides a clean probe
of neutron densities that is free from strong-interaction
uncertainties. By invoking some mild assumptions, PREX provided the
first largely model-independent determination of the neutron radius
$r_{n}$ of ${}^{208}$Pb. Since the charge radius---and its corresponding
proton radius $r_{p}$---is known with enormous
accuracy\,\cite{Angeli:2013}, PREX effectively determined the
neutron-skin of ${}^{208}$Pb to
be\,\cite{Abrahamyan:2012gp}:
$r_{\rm skin}^{208}\!\equiv\!r_{n}\!-\!r_{p}\!=\!{0.33}^{+0.16}_{-0.18}\,{\rm fm}.$
While PREX demonstrated excellent control of systematic errors,
unforeseen technical problems compromised the statistical accuracy
of the measurement. Although such an error is large enough to accommodate 
the predictions of many theoretical models, its large central value of
$r_{\rm skin}^{208}\!=\!{0.33}\,{\rm fm}$ is highly intriguing. It is
intriguing because most nuclear energy density functionals (EDFs)
predict significant lower
values\,\cite{RocaMaza:2011pm,Piekarewicz:2012pp}.

A measurement of the neutron-skin thickness of ${}^{208}$Pb is of
enormous significance due to its very strong correlation to the
slope of the symmetry energy around saturation
density\,\cite{Brown:2000, Furnstahl:2001un, Centelles:2008vu,
RocaMaza:2011pm}. Given that the slope of the symmetry energy $L$ is
presently poorly known, an accurate measurement of $r_{\rm
skin}^{208}$ could help constrain the equation of state (EOS) of
neutron-rich matter, and thus provide vital guidance in areas as
diverse as heavy-ion
collisions\,\cite{Tsang:2004zz,Chen:2004si,Steiner:2005rd,
Shetty:2007zg, Tsang:2008fd} and neutron-star
structure\,\cite{Horowitz:2000xj,Horowitz:2001ya,
Horowitz:2002mb,Carriere:2002bx,Steiner:2004fi,Li:2005sr}.
Conversely, and precisely because of the enormous reach of $r_{\rm
skin}^{208}$, significant constraints on the EOS of neutron-rich
matter are starting to emerge as one combines theoretical,
experimental, and observational
information\,\cite{Piekarewicz:2007dx}. Indeed, a remarkable
consistency seems to appear as one combines laboratory measurements
with astrophysical
observations\,\cite{Tsang:2012se,Lattimer:2012nd}. For example, in
an analysis of the pygmy dipole resonance in exotic nuclei, Carbone
{\sl et al.,} reported values of $L\!=\!(64.8\!\pm\!15.7)$\,MeV\,
and $r_{\rm skin}^{208}\!=\!(0.196\pm0.023)\,{\rm fm}$, finding
remarkable overlap with other methods to extract
$L$\,\cite{Carbone:2010az}. Later on, Steiner and Gandolfi using
predictions from Quantum Monte-Carlo simulations for pure neutron
matter together with neutron-star observations were able to provide
the following stringent limits: $L\!=\!(47.5\!\pm\!4.5)$\,MeV
$\rightarrow$ $r_{\rm skin}^{208}\!=\!(0.171\pm0.007)\,{\rm
fm}$\,\cite{Steiner:2011ft}. Note that the arrow is meant to
indicate that the quoted value of $r_{\rm skin}^{208}$ is derived
from using the strong linear correlation between $L$ and $r_{\rm
skin}^{208}$ obtained in Ref.\,\cite{RocaMaza:2011pm}. By improving
on the finite-range droplet model, M\"oller {\sl et al.} were able
to determine $L\!=\!(70\!\pm\!15)$\,MeV $\rightarrow$
$r_{\rm skin}^{208}\!=\!(0.204\pm0.022)\,{\rm fm}$, although they
recognize that a large variation in $L$ would not significantly alter the
accuracy of their mass model\,\cite{Moller:2012}. Finally, two very recent
compilations have placed constraints on the
density dependence of the symmetry energy from invoking theory,
experiment, and observation\,\cite{Tsang:2012se,Lattimer:2012nd}.
In Ref.\,\cite{Tsang:2012se} (where only theoretical and
experimental information was used) Tsang {\sl et al.,} obtained
constraints of $L\!\sim\!70$\,MeV and $r_{\rm
skin}^{208}\!=\!(0.180\pm0.027)\,{\rm fm}$. Meanwhile, Lattimer
obtained a value of $L\!=\!(50.5\pm9.5)$\,MeV $\rightarrow$ $r_{\rm
skin}^{208}\!=\!(0.175\pm0.014)\,{\rm fm}$\,\cite{Lattimer:2012nd}.
We reiterate that whereas all these predictions for $r_{\rm
skin}^{208}$ can be accommodated comfortably within the PREX
1$\sigma$ error, the PREX central value of $r_{\rm
skin}^{208}\!=\!{0.33}\,{\rm fm}$ is clearly incompatible with all
these findings. Besides these recent analyses, many others have been
published in the literature. However, we are unaware of any analysis
constrained by experimental and observational data that accommodates
a large neutron-skin thickness in ${}^{208}$Pb.

It is the aim of the present contribution to examine critically whether models
with large neutron skins are incompatible with both laboratory and
astrophysical data. To do so we construct new relativistic density functionals
with fairly large values of $r_{\rm skin}^{208}$ that are tested against
existing data. The relativistic EDFs that will be used are based on the interacting
Lagrangian density given in Ref.\,\cite{Todd-Rutel:2005fa}. Such a Lagrangian
density includes a handful of parameters that are calibrated to provide an accurate
description of finite nuclei and a Lorentz covariant extrapolation to dense nuclear
matter. In addition to some of the standard relativistic EDFs used in the literature,
such as NL3\,\cite{Lalazissis:1996rd,Lalazissis:1999}, FSUGold\,\cite{Todd-Rutel:2005fa},
and IU-FSU\,\cite{Fattoyev:2010mx}, we consider three additional EDFs labeled
``TAMUC-FSU'' (or ``TF'' for short) with relatively large neutron skins. Although the
parameters of these models do not follow from a strict optimization procedure, a
significant effort was made in reproducing some bulk parameters of infinite nuclear
matter as well as some critical properties of finite nuclei. However, we note that our
work---devoted exclusively to the study of physical observables---ignores powerful
theoretical constraints that have emerged from the nearly universal behavior of pure 
neutron matter at very low densities. Indeed, the models introduced in this work 
appear inconsistent with such theoretical constraints. Yet, as we show below, it 
seems that such a shortcoming has no impact on the wide range of physical 
observables explored in this work. Further details on the behavior of pure neutron
matter and the calibration procedure will be provided in a forthcoming publication. 

\begin{center}
\begin{table*}[t]
\setlength\extrarowheight{1pt}
\begin{tabular}{|l||c|c|c|c|c|r|r|c|}
 \hline
 \rule{0pt}{.35cm}
 Model & $\rho_{{}_{0}}({\rm fm}^{-3}) $ & $\varepsilon_{{}_{0}}$ & $K_{0}$
          & $\tilde{J}$ & $J$ & $L\hfil$ & $K_{\rm sym}$ &$r_{\rm skin}^{208}$(fm)\\
\hline
\hline
NL3             &  0.148  & $-$16.24 & 271.5 & 25.68 & 37.29 & 118.2 &    100.9 & 0.28  \\
FSU             &  0.148  & $-$16.30 & 230.0 & 26.00 & 32.59 &   60.5 & $-$51.3 & 0.21  \\
IU-FSU          &  0.155  & $-$16.40 & 231.2 & 26.00 & 31.30 &   47.2 &     28.7 & 0.16   \\
TFa             &  0.149  & $-$16.23 & 245.1 & 26.00 & 35.05 &   82.5 & $-$68.4 & 0.25  \\
TFb             &  0.149  & $-$16.40 & 250.1 & 27.59 & 40.07 & 122.5 &      45.8 & 0.30 \\
TFc             &  0.148  & $-$16.46 & 260.5 & 30.20 & 43.67 & 135.2 &      51.6 & 0.33 \\
\hline
\end{tabular}
\caption{Bulk parameters of infinite nuclear matter at saturation density
              $\rho_{_{0}}$ as predicted by the various models used in the
              text. The quantities $\varepsilon_{_{0}}$ and $K_{0}$ represent
              the binding energy per nucleon and incompressibility coefficient
              of symmetric nuclear matter, whereas $J$, $L$, and $K_{\rm sym}$
              denote the energy, slope, and curvature of the symmetry energy
              at $\rho_{_{0}}$; note that $\tilde{J}$ represents the value of the
              symmetry energy at a density of $\rho\!\approx\!0.103\,{\rm fm}^{-3}$. Also
              shown are the predictions for the neutron-skin thickness of ${}^{208}$Pb.
              All quantities are given in MeV unless otherwise indicated.}
\label{Table1}
\end{table*}
\end{center}
Predictions by the six models described in the text for some bulk parameters of symmetric nuclear matter
and of the symmetry energy are displayed in Table\,\ref{Table1}. The notation used for these parameters
follows the convention of Ref.\,\cite{Piekarewicz:2008nh}. Also displayed are the predictions for the
neutron-skin thickness of ${}^{208}$Pb. As advertised, the TF models all predict fairly large
values for $r_{\rm skin}^{208}$, and thus large values for $L$. In what follows we examine whether models
with such large neutron skins are incompatible with available laboratory or astrophysical data.

\begin{figure}[ht]
\smallskip
 \includegraphics[width=7cm,angle=0]{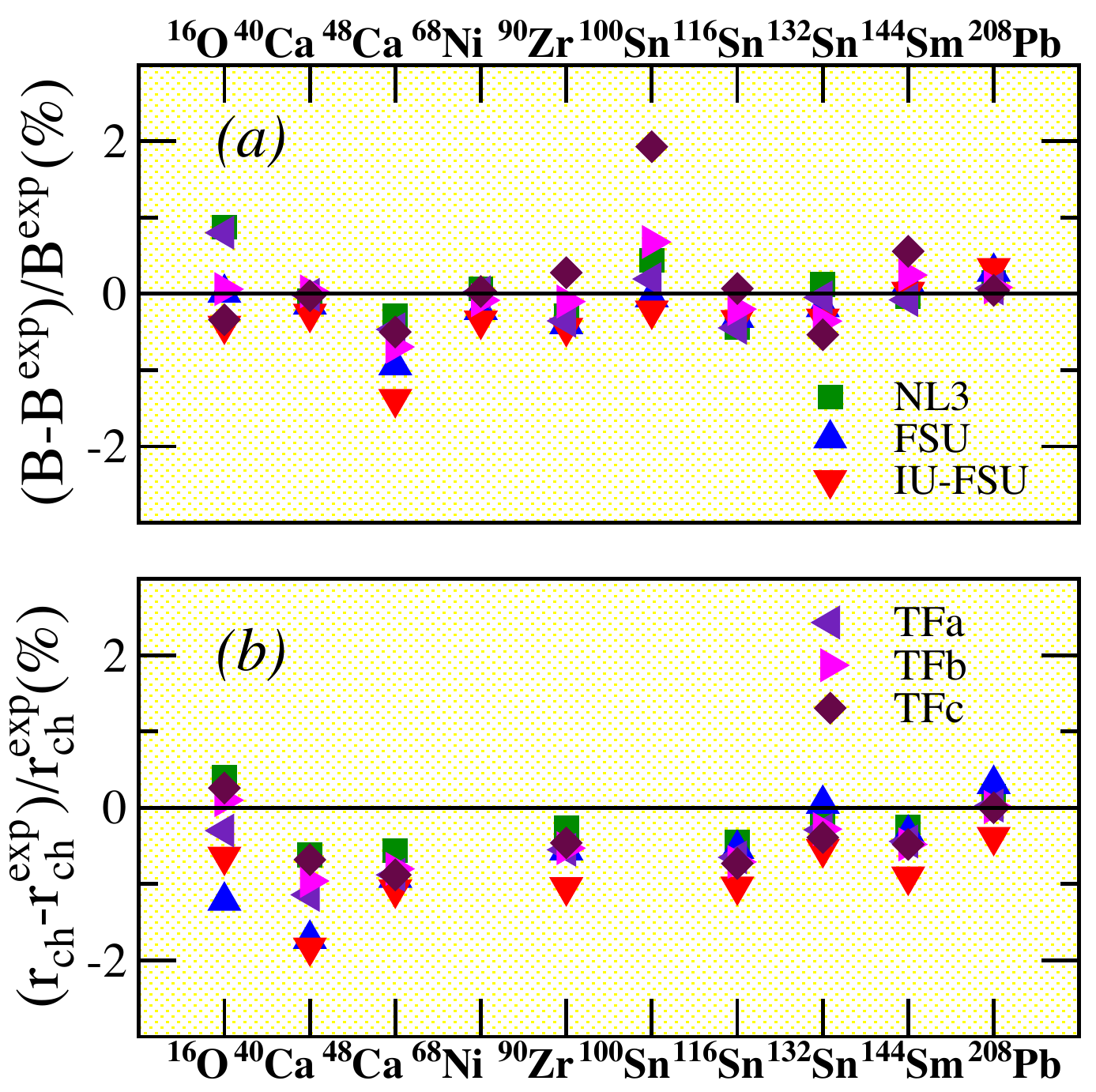}
\caption{(Color online) Residuals (in percentage) using the predictions of the six models
discussed in the text and experiment (when available) for the binding energy (a) and charge
radii (b) of ten magic/semi-magic nuclei across the nuclear chart. Experimental binding
energies were obtained from Ref.\,\cite{Audi:2002rp} and charge radii from
Ref.\,\cite{Angeli:2013}.}
\label{Fig1}
\end{figure}

We start by displaying in Fig.\ref{Fig1} residuals for the binding energy and charge
radius of ten magic or semi-magic nuclei: from ${}^{16}$O to ${}^{208}$Pb. Experimental
data for the binding energies and charge radii were
obtained from Refs.\,\cite{Audi:2002rp} and\,\cite{Angeli:2013}, respectively. Note
that at present there is no available data for the charge radius of neutron-rich
${}^{68}$Ni nor for the neutron-deficient ${}^{100}$Sn\,\cite{Angeli:2013}. One of
the first relativistic EDFs that was accurately calibrated to the ground-state properties
of finite nuclei was NL3\,\cite{Lalazissis:1996rd,Lalazissis:1999}. NL3 has been
enormously successful in reproducing binding energies and charge radii of nuclei
throughout the nuclear chart. However, since the binding energy of stable
nuclei is largely an isoscalar property, the relatively large value of $L$ predicted
by NL3 remained untested. Essentially, nuclear binding  energies are controlled by
the saturation properties of symmetric nuclear matter and the symmetry energy
at a density of about two thirds of that of nuclear matter saturation
(or $\simeq\!0.1\,{\rm fm}^{-3}$)\,\cite{Brown:2000,Furnstahl:2001un}; we denote
this quantity as $\tilde{J}$.
Thus, as illustrated in Fig.\,\ref{Fig1}, NL3 provides a fairly accurate description of
the binding energy and charge radius of all nuclei depicted in the figure. Given
that in all models considered in Table\,\ref{Table1} symmetric nuclear matter
saturates at about the same place and the value of $\tilde{J}$ differs by no more
than $\sim$15\%, we expect an adequate description of binding energies and
charge radii for all the models. This assertion is verified in Fig.\,\ref{Fig1}.
Although only NL3 and FSUGold (or ``FSU'' for short) have been accurately
calibrated, all three TF models provide a description
that is consistent with laboratory data. Given that the values of $L$ tabulated in
Table\,\ref{Table1} vary by more than a factor of two between models, we conclude
that ground-state masses and charge radii are poor isovector indicators that place
no meaningful constraints on the neutron-skin thickness of ${}^{208}$Pb. This
conclusion is in disagreement with the recent findings reported in
Ref.\,\cite{Agrawal:2013hha}.

\begin{figure}[ht]
\smallskip
\includegraphics[width=6cm,height=7cm]{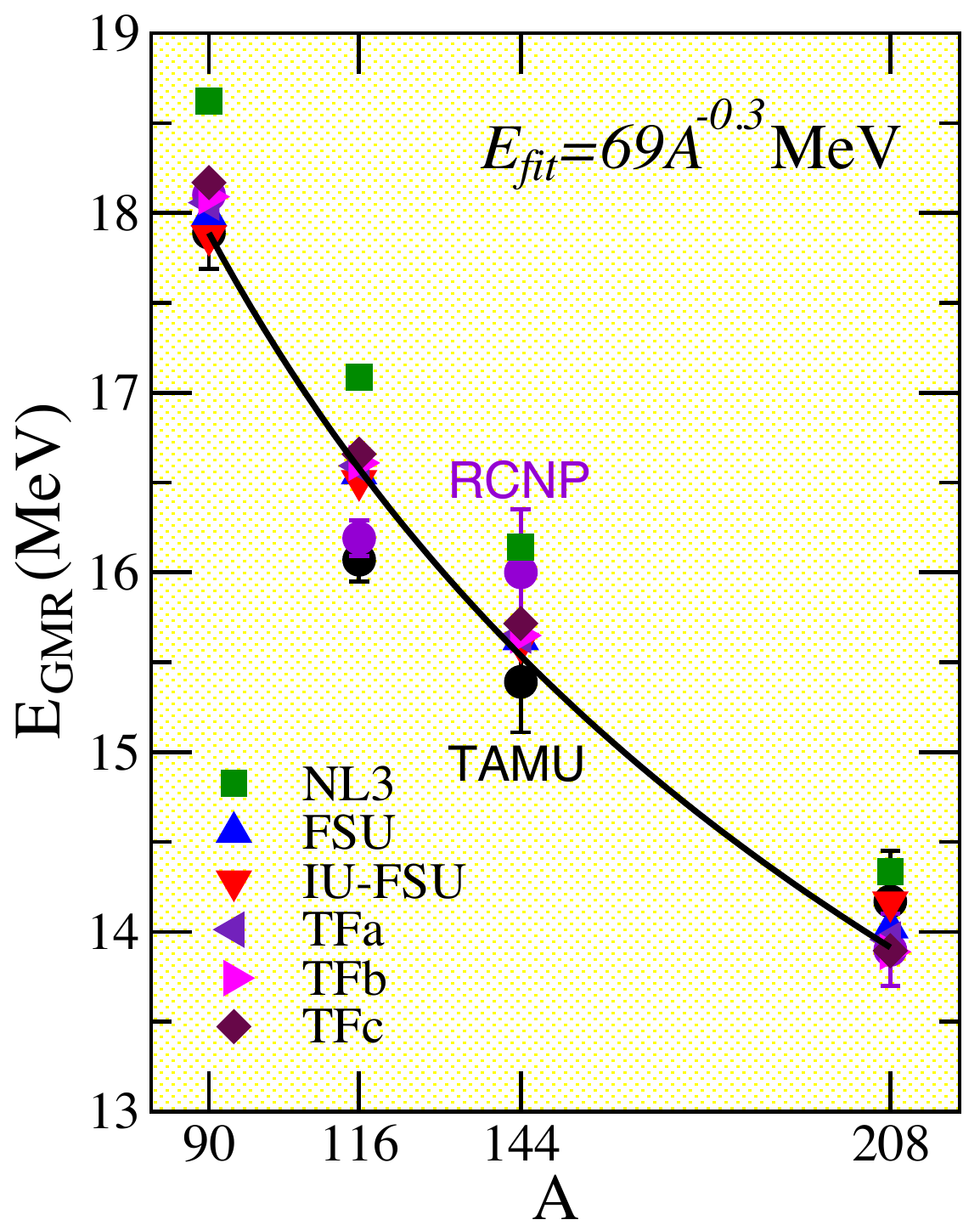}
\caption{(Color online) Predictions for the GMR centroid energies
         of ${}^{90}$Zr, ${}^{116}$Sn, ${}^{144}$Sm, and ${}^{208}$Pb
         from the six models used in the text. Experimental centroid
         energies are from Ref.\,\cite{Youngblood:1999} (TAMU) and
         Refs.\,\cite{Uchida:2003,Uchida:2004bs,Li:2007bp,Li:2010kfa,
         Patel:2013} (RCNP).}
\label{Fig2}
\end{figure}

The collective response of finite nuclei provides a far better test of the isovector
sector than masses and charge radii. In particular, the monopole response (or
{\sl ``breathing mode''}) of neutron-rich nuclei is sensitive to the density dependence
of the symmetry energy. Indeed, the incompressibility coefficient of neutron-rich
matter, a quantity strongly correlated to the breathing-mode energy, may be written as
$K_{0}(\alpha)\!\approx\!K_{0}\!+\!(K_{\rm sym}\!-\!6L\!+\!\ldots)\alpha^{2}$, where
$\alpha\!\equiv\!(N\!-\!Z)/A$ is the neutron-proton asymmetry\,\cite{Piekarewicz:2008nh}.
In Fig.\,\ref{Fig2} we display centroid energies for the giant monopole resonance (GMR)
in ${}^{90}$Zr, ${}^{116}$Sn, ${}^{144}$Sm, and ${}^{208}$Pb\,\cite{Youngblood:1999,
Uchida:2003,Uchida:2004bs,Li:2007bp,Li:2010kfa,Patel:2013}. It is
important to include nuclei with differing values of $\alpha$ since the neutron-proton
asymmetry provides the lever arm for probing the density dependence of the symmetry
energy. For example, whereas NL3---with large  values for both $K_{0}$ and $L$---is
consistent with the measured value of the centroid energy in ${}^{208}$Pb
($\alpha\!=\!0.21$) it overestimates the centroid energy in ${}^{90}$Zr ($\alpha\!=\!0.11$).
The conception of the FSUGold functional was in large part motivated by the desire to
properly describe GMR energies in both ${}^{90}$Zr and ${}^{208}$Pb\,\cite{Piekarewicz:2003br}.
To do so it was required to soften both the EOS of symmetric matter and the symmetry
energy relative to the NL3 predictions\,\cite{Todd-Rutel:2005fa}. Indeed, with such a
softening FSUGold is able to properly describe the experimental GMR energies in all
nuclei, except for the case of ${}^{116}$Sn. Note that the softness of
${}^{116}$Sn in particular---and of all stable Tin isotopes in general---remains an
important open problem\,\cite{Li:2007bp,Piekarewicz:2007us,Piekarewicz:2009gb}.
However, what it  is also evident from the figure is that regardless of the stiffness
of the symmetry energy, all models---with the possible exception of NL3---cluster
around the FSUGold predictions. This suggests that centroid energies of monopole
resonances---even those of nuclei with a large neutron excess---are unable to place
stringent constrains on the neutron-skin thickness of ${}^{208}$Pb.

\begin{figure}[ht]
\smallskip
\includegraphics[width=6cm,height=7cm]{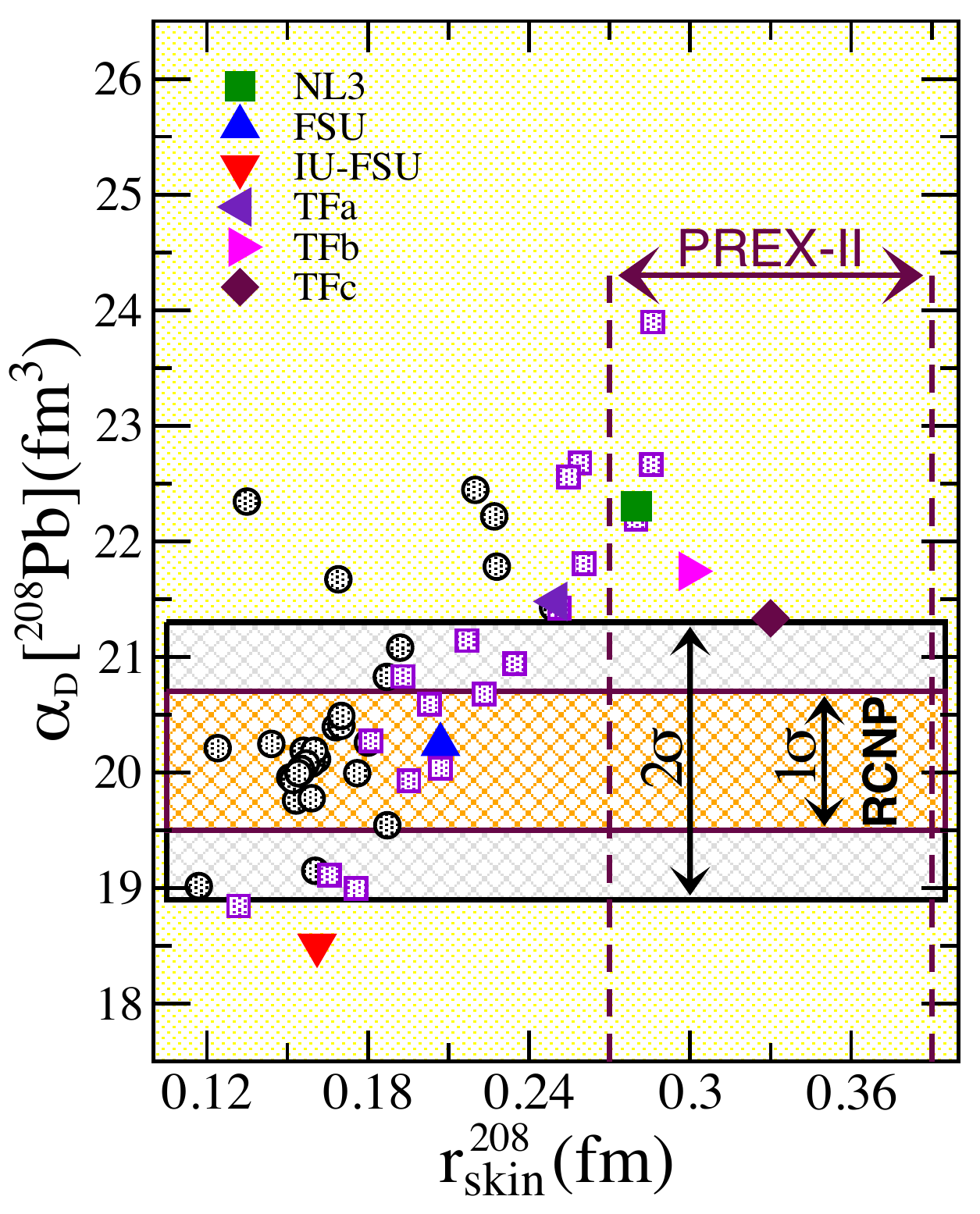}
\caption{(Color online) Predictions from 52 nuclear EDFs for the
electric dipole polarizability and the neutron-skin thickness of
${}^{208}$Pb. Constraints on the dipole polarizability from
RCNP\,\cite{Tamii:2011pv,Poltoratska:2012nf} and from an updated
PREX experiment assuming a 0.06\,fm error and the same central
value\,\cite{Abrahamyan:2012gp} have been
incorporated into the plot.} \label{Fig3}
\end{figure}

In Ref.\,\cite{Reinhard:2010wz} Reinhard and Nazarewicz demonstrated
that the electric dipole polarizability \alphad is a strong
isovector indicator that is strongly correlated to the neutron-skin
thickness of heavy nuclei. Shortly after, using a large number of
EDFs, it was confirmed that such a correlation is
robust\,\cite{Piekarewicz:2012pp}. The electric dipole
polarizability, which is proportional to the inverse energy weighted
sum of the isovector dipole response, is a good isovector indicator
because the symmetry energy acts as the restoring force. The recent
high-resolution measurement of\, \alphad in ${}^{208}$Pb at RCNP
\,\cite{Tamii:2011pv,Poltoratska:2012nf} provides a unique
constraint on $r_{\rm skin}^{208}$ by ruling out models with either
very small ($r_{\rm skin}^{208}\!\lesssim\!0.12$\,fm) or very large
($r_{\rm skin}^{208}\!\gtrsim\!0.24$\,fm) neutron skins. In this
context, the predictions for \alphad from the three stiff TF models
is particularly relevant. To test these models against the RCNP data
we have directly imported the relevant figure from
Ref.\,\cite{Piekarewicz:2012pp}, supplemented with the predictions
from the IU-FSU and TF models. As alluded earlier and clearly
displayed in Fig.\,\ref{Fig3}, the RCNP measurement rules out at the
1$\sigma$ level the predictions of all the models considered in the
text---except for FSUGold. It rules out IU-FSU for having too soft a
symmetry energy, and NL3 and the three TF models for having one that
is too stiff. However, note that the correlation between \alphad and
$r_{\rm skin}^{208}$ is not linear. Indeed, a far better linear
correlation is obtained between $r_{\rm skin}^{208}$ and the product
of $J$\,\alphad\,\cite{Satula:2005hy,Roca-Maza:2013}. Given the
large value of $J$ suggested by the TFc model, this mitigates the
increase of \alphad with $r_{\rm skin}^{208}$, thereby allowing its
prediction to be consistent with the RCNP experiment at the
2$\sigma$ level. Finally, note that we have included the projected
uncertainty of 0.06\,fm for the updated PREX measurement (PREX-II)
assuming that its central value of 0.33\,fm remains intact. If that
proves to be the case, then all 52 models displayed in the figure
will be ruled out!

\begin{figure}[ht]
\smallskip
 \includegraphics[width=7cm,angle=0]{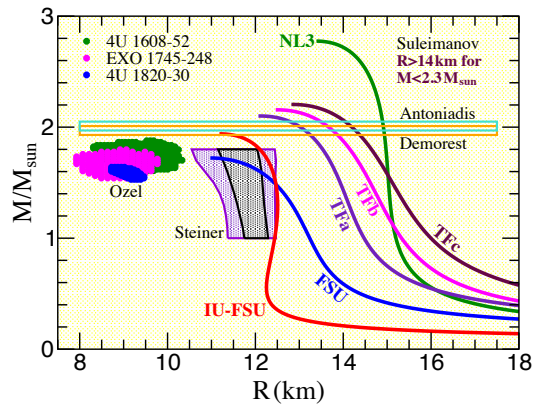}
 \caption{(Color online) {\sl Mass-{\sl vs}-Radius} relation predicted
 by the six models discussed in the text. The horizontal bars in the
 figure represent two accurate measurements of massive neutron
 stars as reported in Refs.\,\cite{Demorest:2010bx,Antoniadis:2013pzd}.
 The mass-radius constraints suggesting very small stellar radii represent
 the 1$\sigma$ confidence contours reported in Ref.\,\cite{Ozel:2010fw}.
 The two shaded areas that suggest larger radii are 1$\sigma$ and 2$\sigma$
 contours extracted from the analysis of Ref.\,\cite{Steiner:2010fz}. Finally,
 the suggestion of large stellar radii by
 Sulemainov {\sl et al.}\,\cite{Suleimanov:2010th} is indicated on the top-right
 hand side of the figure.}
\label{Fig4}
\end{figure}

We finish by displaying in Fig.\,\ref{Fig4} mass-{\sl vs}-radius
relations for neutron stars. Shown with horizontal bars are the two
(accurately measured) massive neutron stars of about 2 solar masses
reported by Demorest {\sl et al.}\,\cite{Demorest:2010bx} and
Antoniadis {\sl et al.}\,\cite{Antoniadis:2013pzd}. Clearly,
theoretical models that predict limiting masses below $2\,M_{\odot}$
(such as FSUGold) require to stiffen the high-density component of
the EOS. Whereas laboratory experiments are of little value in
elucidating the maximum neutron-star mass, they play a critical role
in constraining stellar radii. This is because the same pressure
that pushes against surface tension to create a neutron-rich skin in
nuclei pushes against gravity to determine the size of the neutron
star. Moreover, although neutron stars contain regions of density
significantly higher than those encounter in a nucleus, it has been
shown that stellar radii are controlled by the density dependence of
the symmetry energy in the immediate vicinity of nuclear-matter
saturation density\,\cite{Lattimer:2006xb}. This provides a powerful
{\sl ``data-to-data''}  relation: The larger the neutron-skin
thickness of ${}^{208}$Pb the larger the radius of a neutron
star\,\cite{Horowitz:2001ya}.

Recent advances in X-ray astronomy have allowed for the simultaneous
determination of masses and radii of neutron stars. By reaching this
important milestone one is now able to pose the following
complementary question: how do neutron-star radii constrain the
neutron-skin thickness of ${}^{208}$Pb. The simultaneous
determination of stellar masses and radii has emerged from a study
of three X-ray bursters by \"Ozel and
collaborators\,\cite{Ozel:2010fw}. Results from such a study are
displayed in Fig.\,\ref{Fig4} (in the $8$ to $\!10$ km region) and
suggest very small radii that are difficult to reconcile with the
predictions from all models considered in the text, and indeed from
most models lacking exotic cores\,\cite{Fattoyev:2010rx}. Shortly
after, Steiner, Lattimer, and Brown supplemented \"Ozel's study with
three additional neutron stars and concluded that systematic
uncertainties affect the determination of the most probable masses
and radii\,\cite{Steiner:2010fz}. Their results suggest larger radii
of $11$ to $\!12$ km and have been depicted in Fig.\,\ref{Fig4} by
the two shaded areas that indicate their 1$\sigma$ and 2$\sigma$
contours. It is clear from the figure that the three TF models---as
well as NL3---predict a symmetry energy that is simply too stiff to
be consistent with such an analysis. However, it appears that
systematic uncertainties in the analysis of X-ray bursters continue
to hinder the reliable extraction of stellar radii. Indeed,
Sulemainov and collaborators have suggested that even the more
conservative estimate by Steiner {\sl et al.} must be called into
question\,\cite{Suleimanov:2010th}. The authors of
Ref.~\cite{Suleimanov:2010th} have proposed a lower limit on the
stellar radius of 14\,km for neutron stars with masses below
2.3\,M$_{\odot}$---concluding that neutron-star matter is
characterized by a stiff EOS. Adopting this latest constraint, all
three of the stiff TF models fit comfortably within it. Thus, at present
astrophysical observations are unable to place stringent constraints
on either the density dependence of the symmetry energy or the
neutron-skin thickness of ${}^{208}$Pb.

In summary, we have constructed models with large neutron skins to test whether
the large central value reported by the PREX collaboration may already be ruled
out by existing laboratory or observational data. To do so we have introduced three
models with fairly large neutron skins to compute masses, charge radii, centroid energies
of monopole resonances, the electric dipole polarizability of ${}^{208}$Pb, and
masses and radii of neutron stars. Based on this set of experimental and
observational data, we find no compelling reason to rule out models with large
neutron skins. We should mention, however, that the high-resolution measurement
of the  electric dipole polarizability in ${}^{208}$Pb places a particularly significant
constraint that can only be satisfied at the 2$\sigma$ level. We are confident that
improvements in the statistical and systematic accuracy of future measurements
of neutron skins, dipole polarizabilities, and stellar radii will provide vital constraints
on the density dependence of the symmetry energy. For now, however, ruling out
a thick neutron skin in ${}^{208}$Pb seems premature. 

\begin{acknowledgments}
\vspace{-0.4cm} We would like to thank Profs. Bao-An Li and 
W. G. Newton for fruitful discussions. This work was supported 
in part by grant DE-FD05-92ER40750 from the U.S. Department
of Energy, by the National Aeronautics and Space Administration
under Grant No. NNX11AC41G issued through the Science Mission
Directorate, and the National Science Foundation under Grant No.
PHY-1068022.
\end{acknowledgments}
\vfill

\bibliography{ThickSkins.bbl}
\vfill\eject
\end{document}